\documentclass[aps,prl,twocolumn,amsmath,amssymb,showpacs,
floatfix]{revtex4}
\usepackage{epsfig}
\usepackage{bm}
\begin{document}

%\preprint{}

\title{
%Size dependence of the superconductor-insulator transition in 
%Josephson junction arrays
Entanglement perturbation theory for the quantum ground states in two dimensions
}

%\author{author\altaffilmark{1}, author\altaffilmark{2}, 
%         author\altaffilmark{1,3}}

%\altaffiltext{1}{Affiliation}
%\altaffiltext{2}{Affiliation}
%\altaffiltext{3}{Affiliation}
\author{
S.G. Chung$^{1,2,3}$}
\author{
K. Ueda$^1$}
%\thanks{}
%\altaffiliation
\affiliation{$^1$
Institute for Solid State Physics, the University of Tokyo,
5-1-5 Kashiwanoha, Kashiwa, Chiba 277-8581, Japan\\
$^2$Asia Pacific Center for Theoretical Physics,
Pohang, Gyeonbuk 790-784, Korea\\
$^3$Department of Physics and Nanotechnology Research and Computation Center,
 Western Michigan University, Kalamazoo, MI 49008-5252, USA}
%\affiliation{
%Max-Planck-Institut f\"ur Physik komplexer Systeme, N\"othnitzer Strasse 38, 
%D-01187 Dresden, Germany}

\date{\today}

\begin{abstract}
A simple, general and practically exact method, Entanglement Perturbation Theory (EPT), is formulated to
calculate the ground states of 2D macroscopic quantum systems with 
translational symmetry. An emphasis will be placed on the applicability of 
EPT to fermions.
We will discuss some preliminary evidences
which indicate a potential of EPT.
\end{abstract}

\pacs{71.10.Fd, 71.27.+a, 75.10.Lp}
%\keywords{one dimensional Hubbard model, novel many-body method
%, strongly correlated electrons }

\maketitle

To calculate the partition functions and solve the Schr\"{o}dinger equations for
macroscopic quantum systems are undoubtedly the most fundamental problems in theoretical physics.
It would not be an exaggeration to say that the history of development of quantum theory is primarily
that of pursuit of a variety of methods to this problem. Methods are roughly categorized
into two, a variety of mean field theories and otherwise. The essence of
mean field theories is to make truncations of correlations \cite{kak}. The merit of mean field theories 
lies in an essential simplification of calculations. When the systems of interest are
very complex with many local degrees of freedom, such as involving electrons of
orbital degeneracy and phonons, mean field methods are currently the only methods available.  The penalty is, 
nevertheless, conceptually huge: if we don't know
the importance of those neglected correlations, we are not sure how good the results 
of mean field approximations are. This is precisely the starting point and motivation of non-mean field theories.
Here we currently have roughly three methods. First are rigorous methods such as Bethe Ansatz 
\cite{ons,yan,mac,bax,
and,chu0,lie,sut}.  
In spite of its remarkable success over the last several decades, a major drawback of rigorous methods 
is that they are mostly limited to 2D classical or 1D quantum systems of rather special structures.
Second are numerical simulations such as exact diagonalization and Monte Carlo \cite{suz0}. 
With an ever-increasing computer power and a steady supply of various ingenious ideas in algorithm,
they provide a solid starting point often with a superb accuracy,
but the finite size problem continues to be a serious issue. The negative sign problem is
yet another serious problem in Monte Carlo.  Third is the method of 
NRG (numerical renormalization group), initiated by Wilson \cite{wil}, developed further
first by White's DMRG (density matrix RG) \cite{whi} and most recently by Vidal's 
ER (entanglement renormalization) \cite{vid}.
However, the inability of DMRG for two dimensions is now clear and one still needs to see how much
that inability of DMRG can be improved by ER. Indeed, a deep question remains, namely
 how far one can go
with the very idea of the Hilbert space truncation.

In recent Letters, one of us (SGC) started a novel many-body method to
calculate partition functions for 2,3 dimensional macroscopic classical systems and  
the ground states of macroscopic quantum systems in one dimension \cite{chua,chub}. 
Let us call this method as
EPT (entanglement perturbation theory). We here extend the EPT to 
calculate the quantum ground states in two dimensions.  Of particular interest is
how the EPT handles the  
  fermion anticommutation algebra leading to infinite range correlations among
fermions. Let us call this problem simply the sign 
problem in a wider sense \cite{dag0}.
No existing non-mean-field methods can handle this sign problem satisfactorily.   
We will show that the 2D fermions can be formulated by EPT in a slightly complex
but executable way.
We will give some 
preliminary results for the spin 1/2 Heisenberg antiferromagnet 
and the Hubbard model both on a square lattice, demonstrating
a significant potential of EPT in concept, no RG and no finite size problem, and simplicity 
and practical exactness in 
numerical implementation.
 
\underline{Spins and bosons.} As an example of spins and bosons, let us consider the 
Heisenberg antiferromagnet (HA) on a
square lattice.
The 2D HA model describes the antiferromagnetic interaction between near neighbor spins
with exchange coupling $J$,
\begin{equation}\label{eq18}
H=J\sum_{<ij>}\vec{\sigma_{i}} \cdot \vec{\sigma_{j}}
\end{equation}
where $\vec{\sigma_i}$ is the Pauli spin matrix at the site $i$.
To calculate the ground state of the Schr\"{o}dinger equation
\begin{equation} \label{eq2}
H\Psi=E\Psi
\end{equation}
we follow the following steps. 

{\it First}, instead of (\ref{eq2}), consider the eigenvalue problem for the
density matrix
\begin{equation} \label{eq3}
e^{-\beta H}\Psi=e^{-\beta E}\Psi
\end{equation}
But unlike Monte Carlo and NRG simulations for the ground states, 
our $\beta$ is not the inverse temperature, it is a mere parameter here, $\beta \rightarrow 0$, 
and thus the largest eigenstate of the operator $1-\beta H$ gives the ground state. 
As will become clear below, this simple trick enables EPT to efficiently use the power of
transfer matrix method.  

{\it Second}, we note a decomposition of the density matrix,
\begin{equation} \label{eq7}
e^{-\beta H}=
\Pi_{vc} e^{-\beta H_{vc}}
\Pi_{hc} e^{-\beta H_{hc}}
+\mathcal{O}(\beta^2)
\end{equation}
where "vc" and "hc" denote vertical and horizontal chains, and the every single chain
density matrix is further decomposed into two bond groups: the even group
connecting the sites $(2i,2i+1)$, and the odd group connecting the sites $(2i+1,2i+2)$,
where $i$ is the site index along the chain,
\begin{equation} \label{eq8}
e^{-\beta H_{chain}}
\approx
 e^{-\beta \sum_{even}{H_{bond}}} e^{-\beta\sum_{odd}{H_{bond}}}
\end{equation}
Now the local bond density matrix should be further decomposed as,
\begin{eqnarray} \label{eq9a}
e^{-\beta H_{bond}} 
&\approx&
1-\beta J\vec{\sigma_i} \cdot \vec{\sigma_j}
\nonumber\\
&\equiv&
f_\alpha \otimes g_\alpha
\end{eqnarray}
where and below the repeated indices imply a summation, and $f_\alpha$ takes
four operators, $1$, $\sqrt{\beta J}\sigma_x$, 
$\sqrt{\beta J}\sigma_y$ and $\sqrt{\beta J}\sigma_z$ and $g_\alpha$ likewise 
operators at site $j$.
Thus the matrix product 
representation of the even group bonds in the density matrix is,
$
%\begin{equation} \label{eq11}
\cdots f_{\alpha} \otimes g_{\alpha} \otimes f_{\beta} \otimes g_{\beta} 
\otimes f_{\gamma} \otimes g_{\gamma} \cdots
%\end{equation}
$, and the same expression for the odd group bonds with one lattice shifted from the
even group case.  Putting together, we have the matrix representation of the density matrix
(\ref{eq8}) for a horizontal chain as,
\begin{eqnarray} \label{eq11}
K_{hc} 
&\equiv&
 \cdots \otimes g_{\alpha}\cdot f_{\beta} \otimes f_{\gamma} \cdot g_{\beta}
\otimes g_{\gamma} \cdot f_{\delta} \otimes 
\nonumber\\
&&~~~~~~~~~~f_{\varepsilon} \cdot g_{\delta}
\otimes g_{\varepsilon} \cdot f_{\nu} \otimes \cdots
\nonumber\\
&\equiv& \cdots \Gamma_{\alpha \beta}^1 \otimes \Gamma_{\beta \gamma}^2 \otimes
\Gamma_{\gamma \delta}^1 \otimes \Gamma_{\delta \varepsilon}^2 \cdots
\end{eqnarray}
The density matrix for a vertical 
chain $K_{vc}$ is
 the same as that for a horizontal chain, $K_{hc}$, Eq.(\ref{eq11}). Shifting one lattice
from row to row and column to column, to get a checkerboard-type repetition of $\Gamma^{1,2}$ in 
horizontal and vertical directions, we get a 2D extension of $\Gamma^{1,2}$, namely $\Gamma^1_{\alpha\beta}
({\rm horizontal})\Gamma^2_{\gamma\delta}({\rm vertical}) \rightarrow \Gamma^1_{\alpha\beta\gamma\delta}$,
and $\Gamma^2_{\alpha\beta}
({\rm horizontal})\Gamma^1_{\gamma\delta}({\rm vertical}) \rightarrow \Gamma^2_{\alpha\beta\gamma\delta}$,
where we have used the same notation $\Gamma$ for both 1D and 2D.
The total density matrix in 2D is thus expressed as a bipartite $\Gamma^{1,2}$ network as in Fig.1.
Note that this is a straightforward extension of the 1D case to 2D \cite{chub}.

\begin{figure}
\label{fig1}
\epsfig{file=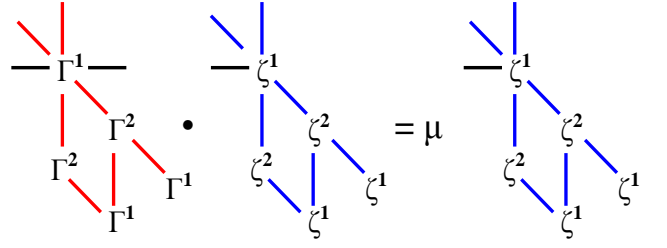,width=8.4cm,height=3.2cm}
\caption{
Schematic representation of the density matrix eigenvalue problem, Eq.(\ref{eq3})
}
\end{figure}
{\it Third}, the ground state wave function can be written also as a straightforward extension of the 
1D case, as in Fig.1, where $\zeta^{1,2}$ now have 4 legs, left-right-up-down each running $1-p$,
where $p$ is the entanglement in 2D, in addition to 
the fifth leg representing the 2 local states $| \uparrow \rangle$ and
$|\downarrow \rangle$.
This tensor product form of the wave function in 2D can be derived by a repeated application
of SVD (singular value decomposition) just like in 1D \cite{chub}. 
Note that the bipartite structure of the total density matrix is a pretense, the Hamiltonian
has a translational symmetry. We choose wave function to be of bipartite structure 
to allow for a possible antiferromagnetic phase. 
We thus arrive at a 
schematic representation, Fig.1, of
the density matrix eigenvalue problem, Eq.(\ref{eq3}). 
An important comment here is on a possibility of broken-symmetry ground states of different types.
For instance, a spin density wave-type ground state may be possible for some
class of systems.  The extension of EPT along this line would be a very 
important future problem \cite{tsu}.

\begin{figure}
\label{fig2}
\epsfig{file=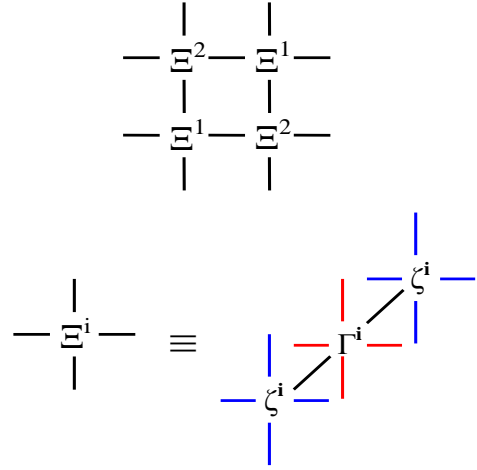,width=6.2cm,height=6.2cm}
\caption{
Schematic representation of the numerator $\Psi K \Psi$.
}
\end{figure}
\begin{figure}
\label{fig3}
\epsfig{file=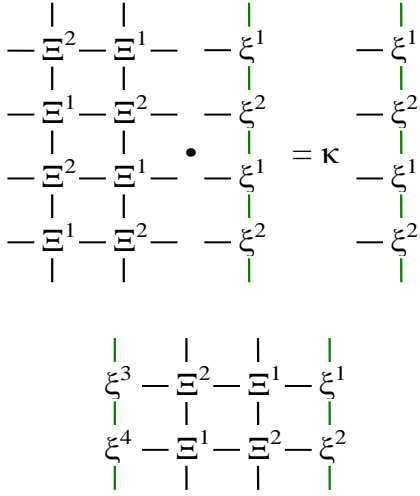,width=5.5cm,height=6.5cm}
\caption{
Schematic representation of a 1D infinite object $\Xi$ and its
eigenvalue problem.  The bottom figure shows a unit structure whose repetition
along the vertical direction forms the quantity $\varphi_L \Xi \varphi_R$. 
}
\end{figure}

{\it Fourth}, we consider the variational problem 
$\mu=\Psi K \Psi/\Psi \Psi \rightarrow max$,
where $K=e^{-\beta H}$,   
by iteration starting with an input
state for $\Psi$. 
The numerator is essentially a partition function of the 2D network of the object 
$\Xi^{1,2}$, cf. Fig.2.  
The 4 legs of the object $\Xi^{1,2}$, 
 runs $1-rp^2$, where $r=4$ is for the interaction channel, cf. Eq.(\ref{eq9a}).
Note that, if $\Xi^1$ and $\Xi^2$ are the same, the calculation is essentially 
that of the partition function for the 2D Ising model \cite{chua}.
Owing to the macroscopic system size, our job is to calculate the 
largest eigenvalue and eigenvector of a 1D infinite object 
shown in Fig.3. Let us denote this object simply as $\Xi$.  Considering
the bipartite structure of $\Xi^{1,2}$ and
absence of right-left symmetry, the appropriate right eigenvector $\varphi_R$ 
should have a bipartite structure, characterized by $\xi^{1,2}$, and likewise
left eigenvector $\varphi_L$ characterized by $\xi^{3,4}$. Again, we solve this eigenvalue
problem by variation.  
We pursue $\varphi_L \Xi \varphi_R/\varphi_L \varphi_R \rightarrow max$.
A local ingredient of the numerator is shown in the bottom of Fig.3.  This quantity is regarded
as a $p^4r^2q^2$ x $p^4r^2q^2$ matrix $A$, where
 $q$ is the second entanglement associated with the wave functions $\varphi_{L,R}$.
The real nonsymmetric matrix $A$ can be written as $A=R\nu L^{tr}$ where the matrices $L$,
$R$, and $\nu$ are made up of left eigenvectors, right eigenvectors and eigenvalues of $A$
and $tr$ means the transpose.  The eigenvectors are normalized as $L^{tr}\cdot R=1$, and 
due to this property, the summation over the combined entanglement-bond indices of dimension
$p^4r^2q^2$ in 
the numerator can be done $N-1$ times, $N \rightarrow \infty$ in the end, and thus we only 
keep the largest eigenvalue $\nu_0$ and eigenvectors ${\bf L_0}$ and 
${\bf R_0}$.  We have $\varphi_L \Xi \varphi_R=\nu_0^{N-1}{\bf L_0}^{tr}
A{\bf R_0}$.
The denominator $\varphi_L \varphi_R$ is handled likewise.
Let us denote the corresponding largest eigenvectors as $\tilde{{\bf L}}_0$ and $\tilde{{\bf R}}_0$ 
and eigenvalue as $\rho_0$.
At this point we realize the "Russian doll" structure, namely the above eigenvalue problems
and our procedures to solve them are essentially those in our study of the 1D model, cf. 
Fig.1 in \cite{chub}.  The $\Xi^{1,2}$ corresponds to $\Gamma^{1,2}$ in the 1D model,
$\xi^{1,2}$ corresponds to $\zeta^{1,2}$ in 1D, with a difference here being the absence 
of left-right symmetry, so that we need the left eigenstate $\varphi_L$ made up of $\xi^{3,4}$
as well as $\varphi_R$ of $\xi^{1,2}$. We will not write down rather lengthy indexes 
involved in the resulting eigenvalue equations for $\xi^{1,3}$ and $\xi^{2,4}$, but
they essentially look like 
\begin{equation}\label{eq13}
M_i(\xi^{1-4})\xi^i=\kappa_0N_i(\xi^{1-4})\xi^i
\end{equation}
for $i=1,2$ where $M_i$ and $N_i$ are square matrices constructed out of $\Xi^{1,2}$, $\xi^{1-4}$, 
${\bf L_0}$, ${\bf R_0}$, $\tilde{{\bf L}}_0$ and $\tilde{{\bf R}}_0$.  
The eigenvalue problem for $\varphi_{L,R}$ is thus reduced to solve these two 
nonlinear, generalized eigenvalue problems by iteration.  Note that the largest eigenvalue 
$\kappa_0$ calculated from the two eigenvalue problems should coincide when the iteration converges.

The above procedure is for $\Psi K \Psi$.  We repeat the procedure for $\Psi \Psi$ which is simply to
replace $\Gamma^{1,2}$ in $\Xi^{1,2}$ by unity, and the resulting object
is denoted as $\Upsilon^{1,2}$. Corresponding to $\varphi_{L,R}$ we write $\chi_{L,R}$,
corresponding to $\xi^{1-4}$ we write $\eta^{1-4}$, and we get similar eigenvalue problems as 
Eq.(\ref{eq13}) for $\eta^{1-4}$.  Technically, we solve the problem for $\eta^{1-4}$ first, and then
use the resulting $\eta^{1-4}$ as a good input for $\xi^{1-4}$, the reason being that the $\Gamma^{1,2}$
are close to unity with a deviation from it an order of $\beta$ which is very small, $10^{-6}$ \cite{chub}.

With the converged $\xi^{1-4}$ and $\eta^{1-4}$, 
we are now ready to derive nonlinear, generalized eigenvalue equations for $\zeta^{1,2}$ 
resulting from the variational problem, $\mu=\Psi K \Psi/\Psi \Psi \rightarrow max$.  
Consider first the numerator.  Its 1D ingredient $\Xi$ can be written as $\Xi=\Phi_L^{tr} \kappa 
\Phi_R$ where $\Phi_{L,R}$ are the square matrices made up of the left, right eigenvectors of $\Xi$,
namely $\varphi_{L,R}$, and $\kappa$ is the diagonal matrix made up of the eigenvalues. Note 
$\Psi K \Psi=T_r(\Xi^N)$, and use the orthogonal property $\Phi_L^{tr}\Phi_R=1$, we then have $\Psi K \Psi
=\kappa_0^{N-1} \varphi_L \Xi \varphi_R$, where the suffix "o" indicates the largest eigenvalue.
 On the other hand, the resulting composite is already analyzed above as,
$\varphi_L \Xi \varphi_R=\nu_0^{N-1}{\bf L_0}^{tr}A{\bf R_0}$. Note that we took in the above variation of the
last quantity with respect to $\xi^{1-4}$.  We now take variation of the same quantity with respect to 
$\zeta^{1,2}$. The two matrices $M_i$ in Eq.(\ref{eq13}) and $X_i$ below are therefore closely related
 to each other.  The denominator $\Psi \Psi$ can be analyzed likewise.  In the end, the variational problem
$\mu=\Psi K \Psi/\Psi \Psi \rightarrow max$ is reduced to nonlinear, generalized eigenvalue problems
for $\zeta^{1,2}$.  It is written schematically as,
\begin{equation}\label{eq14}
X_i(\zeta^{1,2})\zeta^i=\mu_0Y_i(\zeta^{1,2})\zeta^i
\end{equation}
for $i=1,2$, where $X_i$ and $Y_i$ are some symmetric matrices.  We solve this equation for the
next $\zeta^{1,2}$ until convergence. 

{\it Finally} after the convergence, we can calculate various ground state properties.
In general, the expectation value of the two operators 
$\hat{A}$ and $\hat{B}$ sitting on the adjacent sites
is calculated as
$\langle \hat{A}\hat{B} \rangle
=
\Psi\hat{A}\hat{B}\Psi/\Psi\Psi$. 
Now
\begin{eqnarray} \label{eq15}
\Psi\hat{A}\hat{B}\Psi
&=&
\epsilon_0^{N-1}\chi_L\Upsilon (\hat{A}\hat{B})\chi_R
\nonumber\\
&=&
\epsilon_0^{N-1}\lambda_0^{N-1}\acute{{\bf L}}_0^{tr} D(\hat{A}\hat{B})
\acute{{\bf R}}_0
\end{eqnarray}
where $\Upsilon$ is obtained from $\Xi$ by replacing $\Gamma^{1,2} \rightarrow 1$, 
$\chi_{L,R}$ are the left and right eigenvectors of $\Upsilon$ corresponding to 
the largest eigenvalue $\epsilon_0$.  $\Upsilon (\hat{A}\hat{B})$ is obtained from $\Upsilon$ by inserting
$\hat{A}\hat{B}$ into a vertical or horizontal bond.  Likewise the matrix $D$ is obtained from 
the object at the bottom of Fig.3 by replacing $\xi^{1-4}$ by $\eta^{1-4}$ and $\Xi^{1,2}$ by 
$\Upsilon^{1,2}$.  $D(\hat{A}\hat{B})$ is then obtained from $D$ by inserting $\hat{A} \hat{B}$
into a bipartite $\Upsilon^{1,2}$ structure. 
$\lambda_0$, ${\bf \acute{L}_0}$ and ${\bf \acute{R}_0}$ are 
the largest eigenvalue and eigenvectors of $D$.
We thus have
\begin{equation} \label{eq16}
\langle \hat{A}\hat{B} \rangle
=\acute{{\bf L}}_0^{tr}D(\hat{A}\hat{B})\acute{{\bf R}}_0/\lambda_0
\end{equation}

In some preliminary calculations below, we have used the parameter $\beta=10^{-6}$. We have checked 
$\beta=10^{-4}$ with negligible differences.  
The convergence 
criterion is $\| \zeta_{old}^i-\zeta_{new}^i \| / \|\zeta_{old}^i \| \leq 1\cdot10^{-3}$, and for
$\xi$ and $\eta$, the similar criteria be less than $1\cdot10^{-7}$.  
When this condition is met in the latter case, the relative change in the largest eigenvalue  
often hits $10^{-15}$, the machine precision.

\underline{Fermions.} Let us move on to fermions. As a representative
2D fermion system, we consider the Hubbard model on a square lattice,
\begin{equation} \label{eq1}
H=-t\sum_{\sigma,<ij>}(c_{i\sigma}^\dagger c_{j\sigma}+h.c.)
+U\sum_i n_{i\uparrow}n_{i\downarrow}
\end{equation}
where $t$ is the transfer integral, $U$ is the onsite
Coulomb potential and $c_{i\sigma}$, $c_{i\sigma}^\dagger$ are the annihilation
and creation operators for electrons at site $i$ and spin $\sigma$.  We take $t$ as the energy
unit.  
To proceed as in spins and bosons, we 
first rewrite the Hamiltonian (\ref{eq1}) as a sum of a local {\it bond} 
Hamiltonian,
\begin{equation} \label{eq4}
H=\sum_{bond}(H_{ij}+H_i+H_j)\equiv \sum_{bond} H_{bond}
\end{equation}
with
\begin{equation} \label{eq5}
H_{ij}=-t\sum_{\sigma}(c_{i\sigma}^\dagger c_{j\sigma}+h.c.)
\end{equation}
\begin{equation} \label{eq6}
H_i=\frac{U}{4} n_{i\uparrow}n_{i\downarrow}-\frac{\mu}{4}
(n_{i\uparrow}+n_{i\downarrow})
\end{equation}
where the chemical potential 
$\mu$ is introduced to control the electron number per site. 

Following Eqs.(2-5), Eq.(\ref{eq9a}) now becomes 
\begin{eqnarray} \label{eq9}
e^{-\beta H_{bond}} 
%&\approx&
% e^{-\beta H_i} e^{-\beta H_j} e^{-\beta H_{ij}}
%\nonumber\\
&\approx&
 [1-\frac{\beta U}{4} n_{i\uparrow}n_{i\downarrow}+\frac{\beta \mu}{4}
(n_{i\uparrow}+n_{i\downarrow})]\cdot[i \rightarrow j]
\nonumber\\
&    &+ \beta t \sum_{\sigma}(c_{i\sigma}^\dagger c_{j\sigma}+h.c.)
\nonumber\\
&\equiv& \Omega_{\alpha}\otimes \Theta_{\alpha}
\end{eqnarray}
where 
$\Omega_{\alpha}$ takes five operators,  
$1-\frac{\beta U}{4} n_{i\uparrow}n_{i\downarrow}+\frac{\beta \mu}{4}
(n_{i\uparrow}+n_{i\downarrow})$, 
$c_{i\uparrow}^\dagger$, $c_{i\uparrow}$,
$c_{i\downarrow}^\dagger$, and  $c_{i\downarrow}$
and $\Theta_{\alpha}$ likewise operators at site $j$.  

We now come to the point where dimensionality matters.  We need to specify how we represent 
the ground state. We here order the 2D lattice sites as 
\[ \begin{array}{llllllll}
	\cdots & 1  & 2  & 3  & 4  & 5  & 6  & \cdots \\ 
	\cdots & 7  & 8  & 9  & 10 & 11 & 12 & \cdots \\ 
	\cdots & 13 & 14 & 15 & 16 & 17 & 18 & \cdots  
\end{array} \]
where at each lattice site, 
the four basis states are ordered as 
$|0 \rangle$, $| \uparrow \rangle$,
$|\downarrow \rangle$ and $|\uparrow \downarrow \rangle$.  
Let us first consider the density matrix associated with the horizontal chains.
Since the local bond density matrix (\ref{eq9}) contains even number of creation and 
annihilation operators, the matrix representation of the density matrix (\ref{eq9})
reads as 
$
%\begin{equation} \label{eq10}
\cdots
1\otimes 1\otimes \cdots
\langle b_{i+1}| \otimes \langle b_i|
e^{-\beta H_{bond}}|
a_i \rangle \otimes |a_{i+1} \rangle
\cdots
\otimes 1 \otimes 1
\cdots
%\end{equation}
$ where "1" is unit matrix, and therefore
can be written as an operator product of local matrices,
\begin{equation} \label{eq10}
\langle lk|e^{-\beta H_{bond}}|ij \rangle \approx f_{\alpha,ik}
\otimes g_{\alpha,jl}
\end{equation}
where \[ f_1=g_1=\left( \begin{array}{clcr}
			     1 &      0       &     0       &       0 \\
			     0 & \beta \mu /4 &     0       &       0 \\
			     0 &      0       & \beta \mu/4 &       0 \\
			     0 &      0       &     0       & -\beta U/4+\beta \mu/2
			   \end{array} \right)    \]
%$f_2=\sqrt{\beta t}$
 \[ f_2=\sqrt{\beta t}\left( \begin{array}{clcr}
			       0 & 0 & 0 & 0 \\				
			       1 & 0 & 0 & 0 \\				
			       0 & 0 & 0 & 0 \\				
			       0 & 0 & -1 & 0
			       \end{array} \right)  \]
\[ g_2=\sqrt{\beta t}\left( \begin{array}{clcr}
			       0 & 1 & 0 & 0 \\				
			       0 & 0 & 0 & 0 \\				
			       0 & 0 & 0 & 1 \\				
			       0 & 0 & 0 & 0
			       \end{array} \right)  \] etc.
Note that the $-1$ in the $f_2$
matrix is due to the fermion anticommutation algebra.  
With this new $f_{\alpha}$ and $g_{\alpha}$, Eq.(\ref{eq11}) holds for the fermion
$K_{hc}$. Note that due to our choice of ordering of lattice sites, the fermion
anticommutation algebra appears only locally for the horizontal-chain density matrix.

The sign problem, a global effect of anticommutation algebra, arises
for the vertical part.
To see this, consider the matrix element of a local, vertical-bond density matrix
connecting the sites (i,j) and (i+1,j), where row $i$ is one layer above the row $i+1$
and column $j$ is one column left of column $j+1$. The matrix element reads,
\begin{eqnarray} \label{eq12}
\langle b_{i+1,j}| \otimes \langle b_{i+1,j-1}|
\otimes
\cdots
\otimes
\langle b_{i,j+1}| \otimes \langle b_{i,j}|
\nonumber\\
\Omega_{\alpha}(i,j)\otimes \Theta_{\alpha}(i+1,j)
\nonumber\\
|a_{i,j} \rangle 
\otimes
|a_{i,j+1} \rangle
\otimes 
\cdots
\otimes |a_{i+1,j-1} \rangle \otimes 
|a_{i+1,j} \rangle
\nonumber\\
=
f_{\alpha}(i,j) 
\otimes S_{\alpha}(i,j+1)
\otimes S_{\alpha}(i,j+2) 
\cdots
\nonumber\\
\cdots
S_{\alpha}(i+1,j-2) 
\otimes S_{\alpha}(i+1,j-1) 
\otimes 
g_{\alpha}(i+1,j),
\end{eqnarray}
where $S_{\alpha}$ is a unit matrix for the even operator, $\alpha=1$ and 
\[ \left( \begin{array}{cccr}
                               1 & 0 & 0 & 0 \\
                               0 & -1 & 0 & 0 \\
                               0 & 0 & -1 & 0 \\
                               0 & 0 & 0 & 1
                               \end{array} \right) \]
for the odd operators, $\alpha=2-5$.  
\begin{figure}
\label{fig4}
\epsfig{file=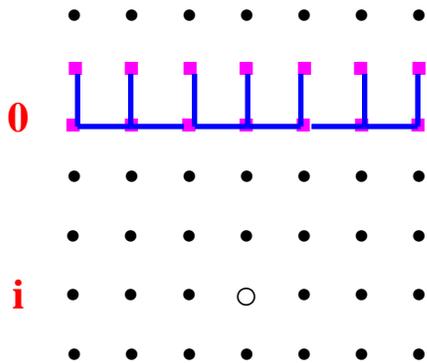,width=5.6cm,height=4.7cm}
\caption{
Taking variation of $\zeta^{1,2}$ in the row $i$, indicated by an open circle.
The unit Hamiltonian is denoted by solid lines. The solid circles
indicate the fermion $\Upsilon^{1,2}$ while the solid rectangles indicate the presence of $S$ matrices.
} 
\end{figure}

The operator network of a bipartite $\Gamma^{1,2}$
with an infinite string of $S$ matrices attached to each 
and every vertical bond, appears hopelessly
complex but not impossible.  
Remember that our parameter
$\beta \rightarrow 0$ in the density matrix.  
Thus we only need to retain the $\beta$ linear terms,
which brings us back to the original Schr\"{o}dinger equation.  
From this viewpoint, we can reformulate EPT in the following way. Variationally,
the Schr\"{o}dinger equation is equivalent to minimize 
$\sum_{unit}\Psi H_{unit} \Psi/\Psi \Psi$ where $H_{unit}$ denotes a 1D subset of the 
Hamiltonian as denoted by solid lines in Fig.4.
Now due to translational symmetry, one can write the numerator as the total number of units times
an expectation value of $H_{unit}$ in the row $0$. Then, a lack of 
translational symmetry of this quantity $\Psi H_{unit} \Psi$ in the vertical direction 
concerning $\zeta^{1,2}$ 
means that we have to variate {\it a vertical string of} $\zeta^{1,2}$,
namely we need to carry out 
\begin{equation} \label{eq17}
\sum_i\delta_i (\Psi H_{unit} \Psi)
\end{equation}
where $\delta_i$ means to variate $\zeta^{1,2}$ in the row $i$, cf. Fig.4. 
Summing over the entire rows $i$, the Hamiltonian and hence the Schr\"{o}dinger equation are
recovered. Clearly, we now have to calculate not only the ground state but also the excited states
 $\chi_{L,R}^k, k=0,1,2,\cdots$ of the fermion version of the infinite object $\Upsilon$ 
and associated eigenvalues. An important note is 
that, due to the translational symmetry of the $H_{unit}$ along the horizontal direction, the 
excited states $\chi_{L,R}^k$ needed here are only those which have the same translational symmetry
as the ground state.  This tremendous simplification is essential in making the EPT analysis
of 2D fermions practically feasible.
In fact, recall that $\Upsilon=\Delta_R\epsilon \Delta_L^{tr}$ where $\Delta_{L,R}$ are the 
left and right eigenvector $\chi_{L,R}$ matrices and $\epsilon$
the diagonal eigenvalue matrix, with orthogonality $\Delta_L^{tr}\Delta_R=1$.
  Inserting this expression for $\Upsilon$, one can easily see that 
the main part of
the calculation in the variation Eq.(\ref{eq17}) is the matrix elements schematically shown
in Fig.5. We calculate $\eta^{1-4}$ as in spins and bosons, but $\xi^{1-4}$ are not
needed here, and the variation of Eq.(\ref{eq17}) with respect to $\zeta^{1,2}$ leads to
similar eigenvalue problems as Eq.(\ref{eq14}).
\begin{figure}
\label{fig5}
\epsfig{file=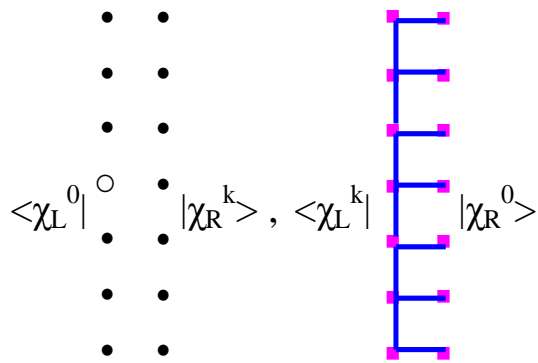,width=4.7cm,height=7.0cm,angle=-90}
\caption{
Schematic representation of the matrix elements which appear in the variation, Eq.(\ref{eq17}).
}
\end{figure}
 
\underline{Preliminary calculations.}
We have done a preliminary calculation on the spin 1/2 HA model,
 cf. Eq(\ref{eq18}). 
The simplest case $p=1$ is soluble by hand.  In fact, putting
\begin{eqnarray} \label{eq19}
\zeta^1={\rm cos}(\alpha) |\uparrow \rangle+{\rm sin}(\alpha) |\downarrow \rangle
\nonumber\\
\zeta^2={\rm cos}(\gamma) |\uparrow \rangle+{\rm sin}(\gamma) |\downarrow \rangle,
\end{eqnarray}
we easily find that the energy per bond is $\frac{1}{4}{\rm cos}[2(\alpha-\gamma)]$
which becomes minimum $-\frac{1}{4}$ at $\alpha-\gamma=\pi/2$, and the magnetic
moments are $\frac{1}{2}{\rm cos}(2\alpha)$ and $-\frac{1}{2}{\rm cos}(2\alpha)$
for the sublattices 1 and 2 corresponding to $\zeta^{1,2}$ in the unit of the Bohr magneton
$\mu_B$.  Thus the ground state at
$p=1$ is the N\'{e}el state with a staggered magnetic moment exactly 1/2, no quantum reduction.
This result was reproduced by our EPT to the machine precision.  Moving on to the case
$p>1$, we have so far examined the case (p,q)=(3,1)
 with a result; energy per bond is $-0.34J$, and the staggered magnetic moment
is 0.39.  
The energy perfectly agrees with the spin wave-Monte Carlo-exact diagonalization
results \cite{man,bea,kat},
 but the staggered moment 0.39 is appreciably larger than their result 0.31,
rather close to the perturbation expansion from the Ising limit to the order of $J^{10}$ \cite{sin},
which gives 0.385.
We certainly need to pursue a stable and 
converging EPT algorithm to reach a clear conclusion.

We have also done a calculation on the 2D Hubbard model ignoring the global, infinite string
of $S$ matrices with an anticipation that this global sign problem should become less and less
important for larger Coulomb repulsion where electrons are strongly localized.  
The calculation takes much more time, and we have tested only one point in the
$U$-electron number per site plane; $U=8$ and half-filling. In the case $(p,q)=(3,2)$,
we have found the ground state energy per site $=-0.473$, and a N\'{e}el state with 
a staggered magnetic moment
0.45.
A remarkable
precision of EPT can be seen in the sublattice magnetic moments,
$\vec{S_1}$ and $\vec{S_2}$.  We found that $\vec{S_1}=- \vec{S_2}$ to
the precision of $10^{-4}$.  In fact, we have chosen the point $U=8$ and
half-filling, because a Monte Carlo result was available for the energy, $-0.48 \pm 0.005$
\cite{hir},
which is very close to our result -0.473. On the other hand, the staggered magnetic moment
appears to be too large, urging again a pursuit of stable and converging EPT algorithm.
An interesting point is that, due to the bipartite
structure, we can check the symmetry of the superconducting order parameter, s-wave or d-wave.  
For instance,
the d-wave order parameter reads as \cite{sca},
\begin{eqnarray} \label{eq20}
\Delta_d=\frac{1}{\sqrt{2}}[(c_{0 \uparrow}c_{x \downarrow}-c_{0 \downarrow}c_{x \uparrow})
-(c_{0 \uparrow}c_{y \downarrow}-c_{0 \downarrow}c_{y \uparrow})
\nonumber\\
+(c_{0 \uparrow}c_{-x \downarrow}-c_{0 \downarrow}c_{-x \uparrow})
-(c_{0 \uparrow}c_{-y \downarrow}-c_{0 \downarrow}c_{-y \uparrow})],
\end{eqnarray}
where $0$ indicates a reference site, $\pm x$ indicates one lattice to the
right or left, and $\pm y$ indicates one lattice up or down.
It should be worth emphasizing that our local basis is grand canonical,
$|0 \rangle$, $| \uparrow \rangle$,
$|\downarrow \rangle$ and $|\uparrow \downarrow \rangle$, and the
electron number per site is controlled by the chemical potential, thus
our EPT has an ability to calculate the superconducting order parameter.  
For $U=8$ and at half-filling with $(p,q)=(3,2)$, 
we have found no superconductivity, the order 
parameter $|\Delta_d| < 4.10^{-4}$, as is anticipated.

In conclusion, we have formulated EPT for the quantum ground states in two dimensions.  The key
point of EPT is that it does not truncate the Hilbert space (no RG), nor
has the finite size problem. Our preliminary calculations on the 2D HA and the Hubbard model
show that EPT is computationally feasible.
A fortunate situation for EPT is that, due to the fact that it is based on 
the local basis,   
the free electron gas is most challenging
but of course soluble by hand.  
Moreover, the fermion sign problem can still be handled by EPT: 
An essential difference from spins and bosons is that we need to calculate the excited states of the
1D infinite object $\Upsilon$ but with an essential simplification that only the
excited states with the same translational symmetry as the ground state need to be
calculated, making the EPT calculation highly executable. 
We thus recognize the following issues for EPT: (1) To establish a stable and converging
numerical procedure.
(2) To calculate the translationally invariant excited 
states of $\Upsilon$ and treat 2D fermions on a rigorous footing.
Finally, it is known that Monte Carlo encounters a negative sign problem not only in fermions but also
in spins on a frustrated lattice such as Kagom\'{e}.  EPT is free from this difficulty, and the
quantum spin liquid issue offers an excellent testing ground for EPT.
\begin{acknowledgments}
This work was partially supported by the NSF under grant No. PHY060010N
and utilized the TeraGrid Cobalt at the National Center for Supercomputing Applications
at the University of Illinois at Urbana-Champaign. SGC thanks Yoshiro Kakehashi, Keisuke Totsuka
and Chitoshi Yasuda for helpful discussions and enlightening comments.
\end{acknowledgments}
\bibliography{q2d}

\begin{thebibliography}{23}
\expandafter\ifx\csname natexlab\endcsname\relax\def\natexlab#1{#1}\fi
\expandafter\ifx\csname bibnamefont\endcsname\relax
  \def\bibnamefont#1{#1}\fi
\expandafter\ifx\csname bibfnamefont\endcsname\relax
  \def\bibfnamefont#1{#1}\fi
\expandafter\ifx\csname citenamefont\endcsname\relax
  \def\citenamefont#1{#1}\fi
\expandafter\ifx\csname url\endcsname\relax
  \def\url#1{\texttt{#1}}\fi
\expandafter\ifx\csname urlprefix\endcsname\relax\def\urlprefix{URL }\fi
\providecommand{\bibinfo}[2]{#2}
\providecommand{\eprint}[2][]{\url{#2}}

\bibitem[{\citenamefont{Y.Kakehashi}(2006)}]{kak}
\bibinfo{author}{\bibnamefont{Y.Kakehashi}}, \bibinfo{journal}{Advances in
  Physics} \textbf{\bibinfo{volume}{53}}, \bibinfo{pages}{497}
  (\bibinfo{year}{2006}).

\bibitem[{\citenamefont{L.Onsager}(1944)}]{ons}
\bibinfo{author}{\bibnamefont{L.Onsager}}, \bibinfo{journal}{Phys.\ Rev.}
  \textbf{\bibinfo{volume}{65}}, \bibinfo{pages}{117} (\bibinfo{year}{1944}).

\bibitem[{\citenamefont{C.N.Yang}(1952)}]{yan}
\bibinfo{author}{\bibnamefont{C.N.Yang}}, \bibinfo{journal}{Phys.\ Rev.}
  \textbf{\bibinfo{volume}{85}}, \bibinfo{pages}{809} (\bibinfo{year}{1952}).

\bibitem[{\citenamefont{B.M.McCoy and T.T.Wu}(1973)}]{mac}
\bibinfo{author}{\bibnamefont{B.M.McCoy}} \bibnamefont{and}
  \bibinfo{author}{\bibnamefont{T.T.Wu}}, \emph{\bibinfo{title}{The Two
  Dimensional Ising Model}} (\bibinfo{publisher}{Harvard University Press},
  \bibinfo{address}{Cambridge, Mass.}, \bibinfo{year}{1973}).

\bibitem[{\citenamefont{R.J.Baxter}(1989)}]{bax}
\bibinfo{author}{\bibnamefont{R.J.Baxter}}, \emph{\bibinfo{title}{Exactly
  Solved Models in Statistical Mechanics}} (\bibinfo{publisher}{Academic
  Press}, \bibinfo{address}{London}, \bibinfo{year}{1989}).

\bibitem[{\citenamefont{N.Andrei et~al.}(1983)\citenamefont{N.Andrei, K.Furuya,
  and J.H.Lowenstein}}]{and}
\bibinfo{author}{\bibnamefont{N.Andrei}},
  \bibinfo{author}{\bibnamefont{K.Furuya}}, \bibnamefont{and}
  \bibinfo{author}{\bibnamefont{J.H.Lowenstein}}, \bibinfo{journal}{Rev.\ Mod.\
  Phys.} \textbf{\bibinfo{volume}{331}} (\bibinfo{year}{1983}).

\bibitem[{\citenamefont{S.G.Chung et~al.}(1983)\citenamefont{S.G.Chung, Y.Oono,
  and Y.C.Chang}}]{chu0}
\bibinfo{author}{\bibnamefont{S.G.Chung}},
  \bibinfo{author}{\bibnamefont{Y.Oono}}, \bibnamefont{and}
  \bibinfo{author}{\bibnamefont{Y.C.Chang}}, \bibinfo{journal}{Phys.\ Rev.\
  Lett.} \textbf{\bibinfo{volume}{51}}, \bibinfo{pages}{241}
  (\bibinfo{year}{1983}).

\bibitem[{\citenamefont{E.H.Lieb and F.Y.Wu}(1968)}]{lie}
\bibinfo{author}{\bibnamefont{E.H.Lieb}} \bibnamefont{and}
  \bibinfo{author}{\bibnamefont{F.Y.Wu}}, \bibinfo{journal}{Phys.\ Rev.\ Lett.}
  \textbf{\bibinfo{volume}{25}}, \bibinfo{pages}{1445} (\bibinfo{year}{1968}).

\bibitem[{\citenamefont{B.Sutherland}(2004)}]{sut}
\bibinfo{author}{\bibnamefont{B.Sutherland}}, \emph{\bibinfo{title}{Beautiful
  models: 70 years of exactly solved quantum many-body problmes}}
  (\bibinfo{publisher}{World Scientific}, \bibinfo{address}{New Jersey},
  \bibinfo{year}{2004}).

\bibitem[{\citenamefont{M.Suzuki}(1993)}]{suz0}
\bibinfo{editor}{\bibnamefont{M.Suzuki}}, ed., \emph{\bibinfo{title}{Quantum
  Monte Carlo methods in condensed matter physics}} (\bibinfo{publisher}{World
  Scientific}, \bibinfo{address}{Hong Kong}, \bibinfo{year}{1993}).

\bibitem[{\citenamefont{K.G.Wilson}(1975)}]{wil}
\bibinfo{author}{\bibnamefont{K.G.Wilson}}, \bibinfo{journal}{Rev.\ Mod.\
  Phys.} \textbf{\bibinfo{volume}{47}}, \bibinfo{pages}{773}
  (\bibinfo{year}{1975}).

\bibitem[{\citenamefont{S.R.White}(1993)}]{whi}
\bibinfo{author}{\bibnamefont{S.R.White}}, \bibinfo{journal}{Phys.\ Rev. B}
  \textbf{\bibinfo{volume}{48}}, \bibinfo{pages}{10345} (\bibinfo{year}{1993}).

\bibitem[{\citenamefont{G.Vidal}(2007)}]{vid}
\bibinfo{author}{\bibnamefont{G.Vidal}}, \bibinfo{journal}{Phys.\ Rev.\ Lett.}
  \textbf{\bibinfo{volume}{99}}, \bibinfo{pages}{220405}
  (\bibinfo{year}{2007}).

\bibitem[{\citenamefont{S.G.Chung}(2006)}]{chua}
\bibinfo{author}{\bibnamefont{S.G.Chung}}, \bibinfo{journal}{Phys.\ Lett. A}
  \textbf{\bibinfo{volume}{359}}, \bibinfo{pages}{707} (\bibinfo{year}{2006}).

\bibitem[{\citenamefont{S.G.Chung}(2007)}]{chub}
\bibinfo{author}{\bibnamefont{S.G.Chung}}, \bibinfo{journal}{Phys.\ Lett. A}
  \textbf{\bibinfo{volume}{361}}, \bibinfo{pages}{396} (\bibinfo{year}{2007}).

\bibitem[{\citenamefont{E.Daggato}(1994)}]{dag0}
\bibinfo{author}{\bibnamefont{E.Daggato}}, \bibinfo{journal}{Rev.\ Mod.\ Phys.}
  \textbf{\bibinfo{volume}{66}}, \bibinfo{pages}{763} (\bibinfo{year}{1994}).

\bibitem[{\citenamefont{H.Tsunetsugu}(private communication)}]{tsu}
\bibinfo{author}{\bibnamefont{H.Tsunetsugu}} (\bibinfo{year}{private
  communication}).

\bibitem[{\citenamefont{E.Manousakis}(1991)}]{man}
\bibinfo{author}{\bibnamefont{E.Manousakis}}, \bibinfo{journal}{Rev.\ Mod.\
  Phys.} \textbf{\bibinfo{volume}{63}}, \bibinfo{pages}{1}
  (\bibinfo{year}{1991}).

\bibitem[{\citenamefont{B.B.Beard and U.-J.Wiese}(1996)}]{bea}
\bibinfo{author}{\bibnamefont{B.B.Beard}} \bibnamefont{and}
  \bibinfo{author}{\bibnamefont{U.-J.Wiese}}, \bibinfo{journal}{Phys.\ Rev.\
  Lett.} \textbf{\bibinfo{volume}{77}}, \bibinfo{pages}{5130}
  (\bibinfo{year}{1996}).

\bibitem[{\citenamefont{K.Kato et~al.}(2000)\citenamefont{K.Kato, S.Todo,
  K.Harada, N.Kawashima, S.Miyashita, and H.Takayama}}]{kat}
\bibinfo{author}{\bibnamefont{K.Kato}}, \bibinfo{author}{\bibnamefont{S.Todo}},
  \bibinfo{author}{\bibnamefont{K.Harada}},
  \bibinfo{author}{\bibnamefont{N.Kawashima}},
  \bibinfo{author}{\bibnamefont{S.Miyashita}}, \bibnamefont{and}
  \bibinfo{author}{\bibnamefont{H.Takayama}}, \bibinfo{journal}{Phys.\ Rev.\
  Lett.} \textbf{\bibinfo{volume}{84}}, \bibinfo{pages}{4204}
  (\bibinfo{year}{2000}).

\bibitem[{\citenamefont{R.R.P.Singh}(1989)}]{sin}
\bibinfo{author}{\bibnamefont{R.R.P.Singh}}, \bibinfo{journal}{Phys.\ Rev. B}
  \textbf{\bibinfo{volume}{39}}, \bibinfo{pages}{9760} (\bibinfo{year}{1989}).

\bibitem[{\citenamefont{J.E.Hirsch}(1985)}]{hir}
\bibinfo{author}{\bibnamefont{J.E.Hirsch}}, \bibinfo{journal}{Phys.\ Rev. B}
  \textbf{\bibinfo{volume}{31}}, \bibinfo{pages}{4403} (\bibinfo{year}{1985}).

\bibitem[{\citenamefont{D.J.Scalapino}(1995)}]{sca}
\bibinfo{author}{\bibnamefont{D.J.Scalapino}}, \bibinfo{journal}{Physics
  Report} \textbf{\bibinfo{volume}{250}}, \bibinfo{pages}{329}
  (\bibinfo{year}{1995}).

\end{thebibliography}

\end{document}